\documentclass{PoS}

\title{Standard Model $H \rightarrow \gamma \gamma$ discovery potential with ATLAS}

\ShortTitle{Standard Model $H \rightarrow \gamma \gamma$ discovery potential with ATLAS}

\author{\speaker{Yaquan FANG}\\%
        On Behalf Of The ATLAS Collaboration \\
        University of Wisconsin, Madison\\
        E-mail: \email{yaquan.fang@cern.ch}}

\abstract{This contribution summarizes the discovery potential of the
  Standard Model Higgs boson using the $H\rightarrow \gamma \gamma$
  decay with the ATLAS detector. The relevant detector performance
  aspects of photon reconstruction, photon identification and trigger
  issues are discussed. The potential of inclusive $H\rightarrow
  \gamma \gamma$ as well as Higgs boson searches in association with
  one or two hard jets are studied. %The associated
  The discovery potential is finally assessed using an unbinned
  multivariate maximum-likelihood fit with an expected integrated
  luminosity of $\approx 10 fb^{-1}$.}

\FullConference{Physics at LHC 2008\\
		 29 September - October 4, 2008\\
		 Split, Croatia}

\begin{document}

\section{Introduction}
In the light Higgs mass range $115<m_{H}<140$ GeV, $H\rightarrow
\gamma \gamma$ is one of the most promising channels. Excellent photon energy
reconstruction and angular resolution are required to observe a
narrow diphoton mass peak above the $\gamma\gamma$ QCD invariant mass
spectrum. Furthermore, an effective photon identification is necessary
to reject the huge rate of QCD jets and $\gamma$-jets. This channel has been studied for a long
time~\cite{atlas1}. The inclusive study is revisited with the most up-to-date detector
geometry and software. The impact of higher order QCD and EW
corrections is also considered in this study (see Ref.~\cite{csc}). The analysis of Higgs
decaying into two photons in association with one or two hard jets is
re-evaluated with a more complete description of the detector
simulation. The analyses corresponding to associated processes $WH,
ZH$ and $t \overline{t}$ are also updated and documented in Ref~\cite{csc}. In addition to the invariant mass of the photon pair, other
kinematics and different topological properties are also taken into account to enhance the
sensitivity of the analysis. All the aspects are incorporated by means of a simultaneously
unbinned maximum-likelihood fit. 

\section{Experimental aspects of the analysis}
\subsection{Trigger items}
Two trigger menus 2g17i and g55 are expected for $H\rightarrow \gamma\gamma$
analysis. The former selects events with two isolated photons with
$p_{T}>20$ GeV. At least one photon with $p_{T}>60$ GeV is required
for selected events in the g55 trigger. In this study, 2g17i is used to avoid
the biases in the diphoton mass reconstruction. The signal's trigger
efficiency, normalized to inclusive analysis cuts, is $94\%$.

\subsection{Photon identification and jet rejection}
Powerful photon identification is crucial for a search in the $H\rightarrow
\gamma\gamma$ channel. A single jet rejection with a factor of 5000 is
required while keeping the high transverse momentum photons ($p_{T}>
25$ GeV) with an efficiency of $\sim 80\%$. 
Dedicated cuts~\cite{csc} were applied on shower shape variables based on EM calorimeter depositions and track isolation. The rejection for an inclusive jet with track isolation cut is
8160$\pm$250 (5070$\pm$120 without track isolation cut). The rejection
against gluon initiated jets is a factor of 10 higher than
that of quark initiated jets, mostly due
 to the different fragmentation characteristics. The average EM
 calorimeter cut efficiency for signal sample has been found to be $\sim83\%$, even in
 presence of pile-up. The track isolation cut efficiency is $\sim98\%$.

\subsection{Conversion reconstruction}
Converted photons showering upstream can result in the energy resolution worse than
that of the unconverted photons. In addition, their energy deposition in the
calorimeter is geometrically wider in the $\phi$
direction due to the magnetic field in the Inner Detector. Nevertheless,
the track of the converted $e^{+}/e^{-}$ can help
to improve the determination of the primary vertex.  
Photon conversion needs to be taken into account carefully since
around $\sim57\%$ of selected events have at least one converted
photon. A vertexing algorithm using a reconstructed particle track has been
developed to associate an EM cluster (converted photon candidate) with two or one tracks, namely
double track conversion or single track conversion. The tagging
efficiency as a function of conversion is shown in Figure~\ref{fig:photon_conversion}.

\subsection{Primary vertex reconstruction}

\begin{figure}[!h]
\begin{minipage}[b]{.46\linewidth}
\centering\includegraphics[scale=0.39]{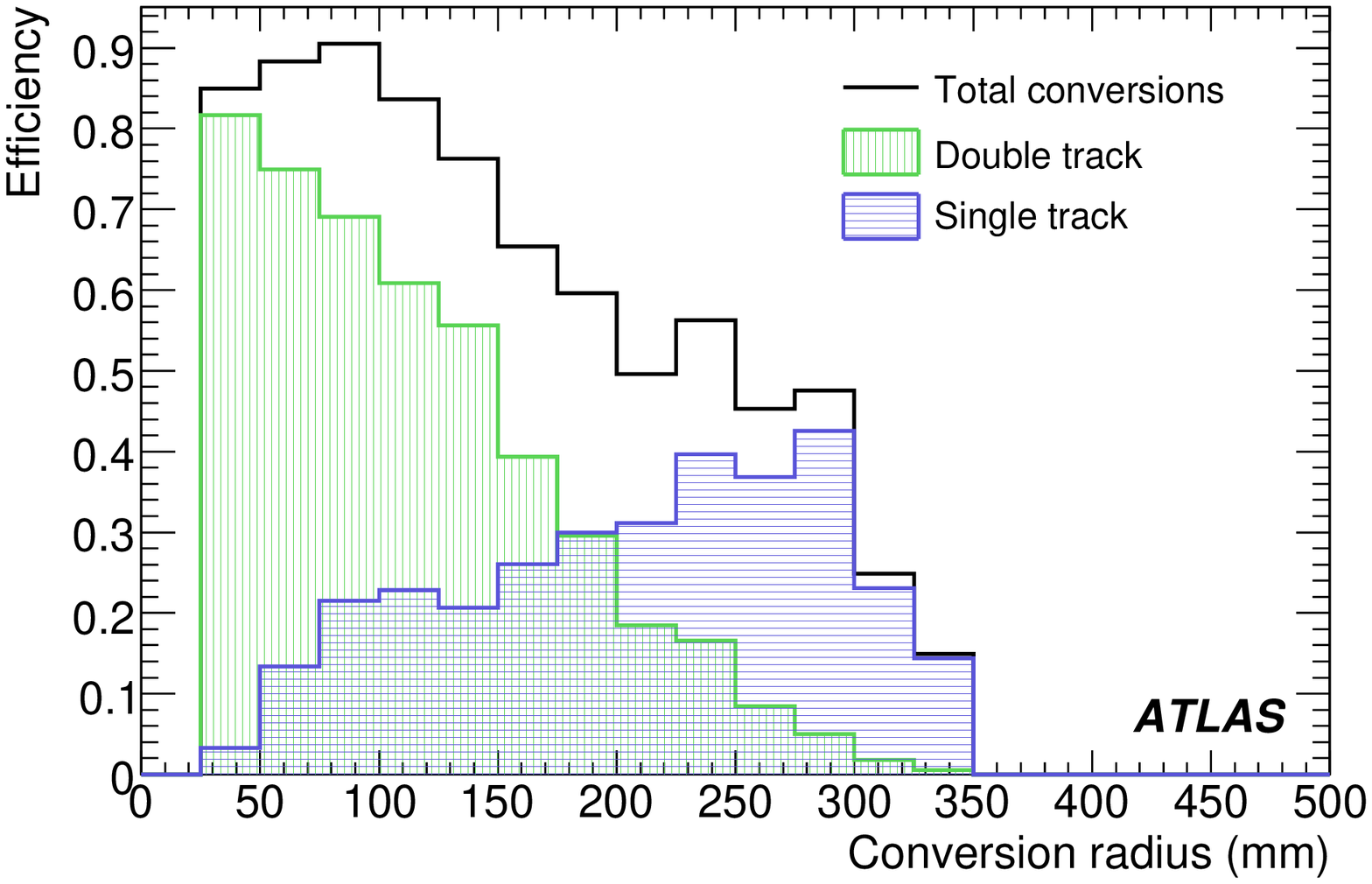}
\caption{Efficiency of single and double track conversion
  reconstruction as a function of conversion radius.}
\label{fig:photon_conversion}
\end{minipage} \hfill
\begin{minipage}[b]{.46\linewidth}
\centering\includegraphics[scale=0.28]{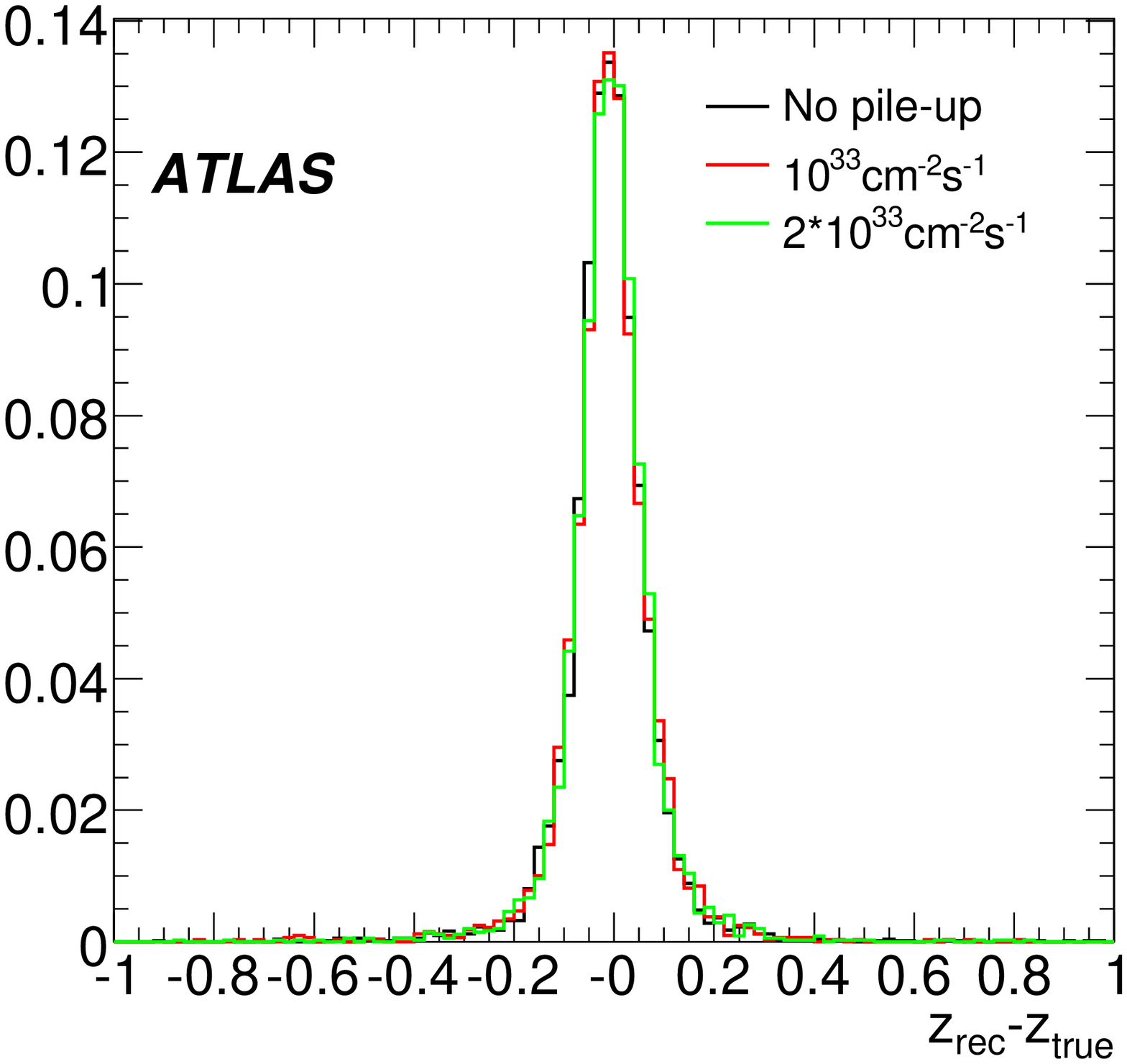}
\caption{Calorimetric pointing with the
  reconstructed primary vertex. }
\label{fig:photon_vertex}
\end{minipage}
\end{figure}

Precise measurement of $z$ vertex of the two photons can help to determine the $\eta$ of the photons,
thereby enhancing the mass resolution of the reconstructed Higgs. To do
this, linear fits
are implemented in the $(R,z)$ plane through the cluster barycenters detected in the
presampler and in the first and second samplings of the EM
calorimeter. The intercept of the lines with the beam axis provides the
hard scattering vertex of the photons. Combining them and the nominal
interaction vertex at $z_{0} = 0$ yields an estimation of $z$ vertex of Higgs boson based on
measurements in the calorimeter, $Z_{H}^{calo}$. In the case of conversion, the
Higgs boson vertex position accuracy can be improved by including the
pointing of the conversion track. By adding the reconstructed event
vertex to the linear fit, the best Higgs boson position accuracy is
achieved with a Gaussian width of 0.07 mm as
Figure~\ref{fig:photon_vertex} shows. The discrimination of the hard-scattering
vertex from those due to pile-up is done using a likelihood method.

%A tagged converted photon with
%transverse energy in the EM calorimeter $E_{T}$ has some tracks associated with it,
%providing a measurement of the transverse momentum of the converted
%photon in tracker, say $p_{T}$. The distribution of $p_{T}/E_{T}$
%exhibit distinction between converted photon and converted photon from
%jet ($\pi^{0}$) as right plot of Figure~\ref{fig:photon_conversion}
%shows. The different shapes can be parametrized and used to
%discriminate between components in a data sample of conversion events,
%allowing the evaluation of the $\gamma j$ and $jj$ percentage in the background.

\subsection{Invariant mass reconstruction}
\begin{figure}[!h]
\begin{minipage}[b]{.46\linewidth}
\centering\includegraphics[scale=0.38]{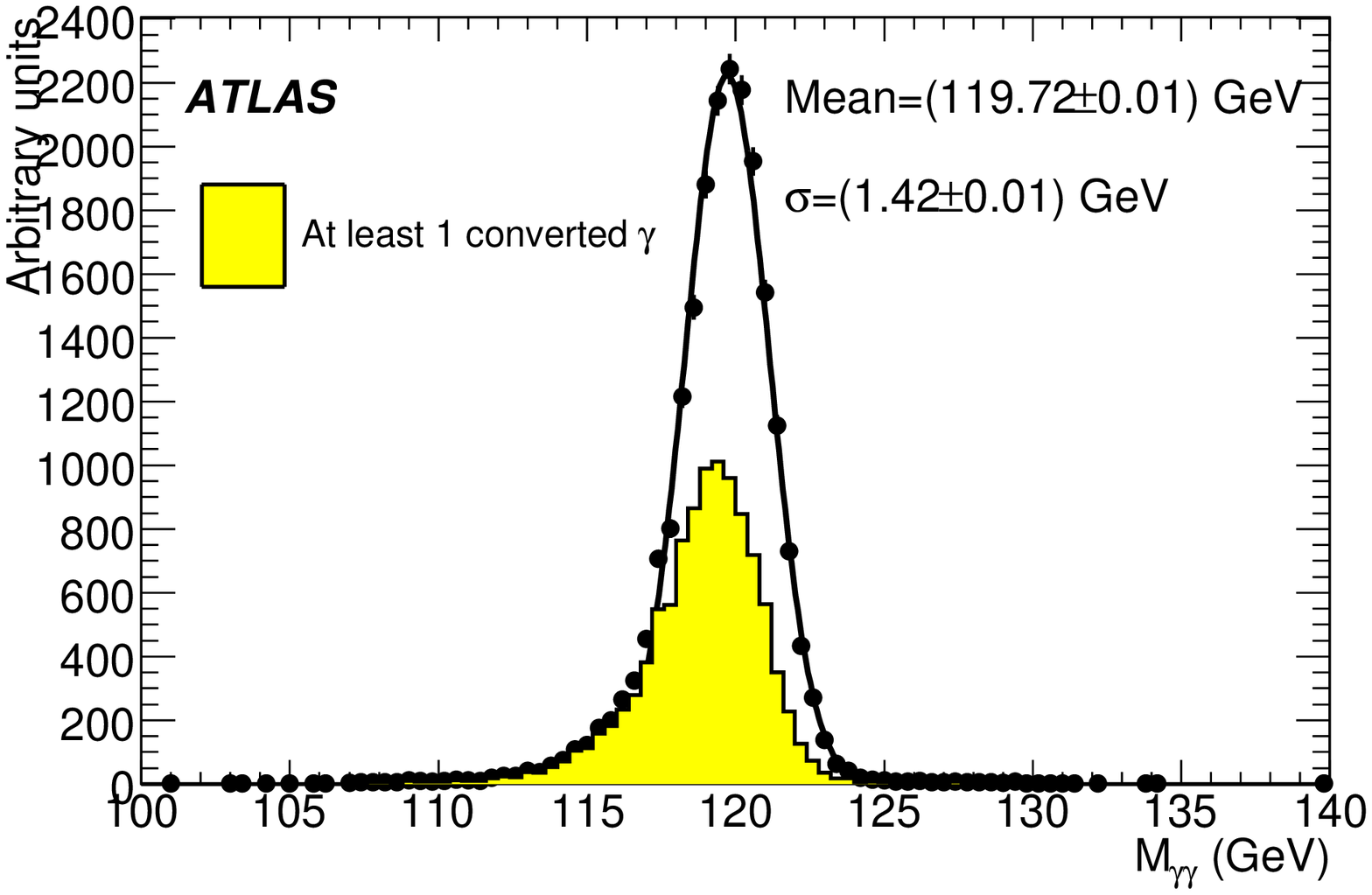}
\caption{Invariant mass distributions for diphotons from Higgs boson
  decays with $m_{H} = 120$ GeV. The shaded histogram corresponds to events with at least one converted photon.}
\label{fig:massreco}
\end{minipage} \hfill
\begin{minipage}[b]{.46\linewidth}
\centering\includegraphics[scale=0.4]{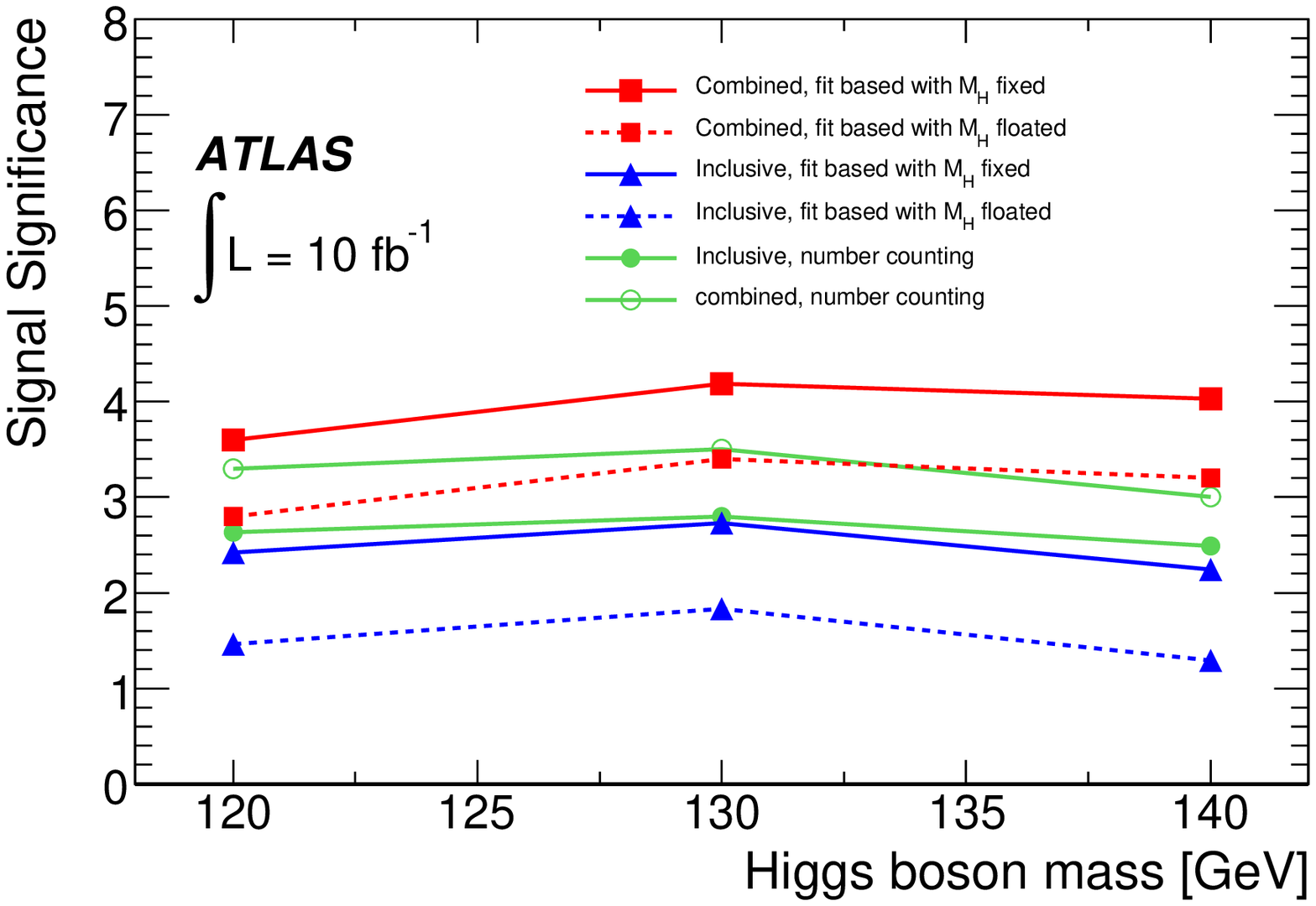}
\caption{Expected signal significance for a Higgs boson using
  the $H\rightarrow\gamma\gamma$ decay for $10\,$fb$^{-1}$ of
  integrated luminosity as a function of the mass (see text for more explanation). }
\label{fig:significance}
\end{minipage} \hfill
\end{figure}

Unconverted photons and converted photons are reconstructed from
clusters of different size. The energy of the photons in the EM
calorimeter has been reconstructed and calibrated considering a series
of effects. The invariant mass of the diphotons has been
reconstructed after the trigger and inclusive
cuts. Figure~\ref{fig:massreco} presents the invariant mass distributions for diphotons from Higgs boson
decays with $m_{H} = 120$ GeV. The mass resolution has been
determined from an asymmetric Gaussian fit $([-2\sigma,+3\sigma])$ to
the invariant mass peak. The relative mass resolution
$\sigma_{m}/m$ is close to  1.2$\%$ degrading by a few percent when pile-up
is added.

\section{Event Selection}
\label{sec:eventsel}
\subsection{Inclusive analysis}
The signature for the inclusive analysis is simply two energetic photons. 
We require that two photon candidates in the central detector region, defined as
$\left|\eta\right|<2.37$, excluding the crack region
between barrel and end-cap calorimeters,
$1.37<\left|\eta\right|<1.52$.  At this level it is required that the events pass the trigger selection.
Transverse momentum cuts of 40 and 25 GeV are applied on the leading and sub-leading photon candidates, respectively.

\subsection{Higgs boson plus one jet and Higgs boson plus two jets
  analysis}
Photonic cuts for the Higgs boson plus one jet analysis are similar to 
those of the inclusive case, except that the $p_{T}$ cut on leading photon is 45
GeV instead of 40 GeV. In addition, we require at least one hadronic jet with
$p_{T}>20$ GeV in $|\eta|<5$. The invariant mass of the diphotons and
the leading jet ($m_{\gamma\gamma j}$) is used as a discriminator to
improve the signal-to-background ratio. The cut on it is
$m_{\gamma\gamma j}>350$ GeV.

The transverse momentum cut on the leading photon for the Higgs boson
associated with two jets is 5 GeV harder than
that for Higgs boson plus one jet analysis. In addition, at least two
hadronic jets are required in $|\eta|<5$ with $p_{T}>40,20$ GeV for the leading and
sub-leading jet, respectively. These two tagging jets are required to
be in opposite hemispheres and pseudo-rapidity gap between them should
be larger than 3.6. Photons are required to have pseudorapidity
between those of the tagging jets. The invariant mass of the tagging
jets $m_{jj}$ is required to be greater than 500 GeV. Finally, we veto
any event with a third jet which has $p_{T}>20$ GeV and $|\eta|<3.2$.

%\subsection{Higgs+isolated leptons+$E_{T}^{miss}$ and Higgs +$E_{T}^{miss}$ }

%The main signal processes contribute to $H+E_{T}^{miss}+1\ell$ and $H+E_{T}^{miss}$ are $WH\rightarrow \mu \gamma
%\gamma$, $t \overline{t} H$ and $ZH\rightarrow \nu \nu \gamma \gamma$ respectively. The cuts applied on two photons are
%with $p_{T}>60,30$ GeV in both cases. At least one energetic isolated lepton with $p_{T}>30$ GeV is
%required and the cut on $E_{T}^{miss}$ is 30 GeV in the
%$H+E_{T}^{miss}+1\ell$. $H+E_{T}^{miss}$ channel has $E_{T}^{miss}$
%cut as high as 80 GeV. There are also some cuts to remove  $w\gamma$
%and $t \overline{t} \gamma \gamma$ and  $b \overline{b} \gamma \gamma$
%and results are documented in Ref~\cite{csc}.

\section{ATLAS Higgs discovery potential in the $\gamma \gamma$ channel}

This section reports the discovery potential of a Higgs boson in the mass range $120<
m_{H}<140$ GeV. In inclusive analysis case, the expected signal significances using event counting for 10 fb$^{-1}$ of
integrated luminosity are 2.6, 2.8 and 2.5 with respect to Higgs boson
masses of 120, 130 and 140 GeV. Since the sensitivity for the inclusive,
H+1jet and H+2jets selections are different, a combination of them
taking into account the event overlap among the three analyses, can
enhance the significance to 3.3 (120 GeV), 3.5 (130 GeV), 3.0 (140
GeV).

In parallel, the significances are computed using the maximum-likelihood fit formalism. With respect to the cut analyses, the fit procedure
takes advantage of further discrimination information from the kinematic
and topological properties of $H\rightarrow \gamma \gamma$ decays. In
addition to the diphoton mass, $m_{\gamma\gamma}$, the transverse momentum
of the Higgs boson, $P_{T,H}$, and the magnitude of the photon decay
angle in the Higgs boson rest frame with respect to the Higgs laboratory
flight direction, $|cos(\theta^{*})|$, are included. Three photon pseudorapidity regions according to
the resolution of $m_{\gamma\gamma}$ and three jet categories as well
as photon conversion are introduced to classify events during the fit
processes. The significances are extrapolated in both cases of fixed
and floating Higgs mass.
Figure~\ref{fig:significance} displays a summary of the expected
signal significance for the inclusive and final combined analysis for
$10\,$ fb$^{-1}$ of integrated luminosity as a function of the Higgs
boson mass. The solid and hollow circles
  correspond to the sensitivity of the inclusive and combined analysis using event
  counting. The solid triangles linked with solid and dashed
  lines correspond to the sensitivity of the inclusive analysis by
  means of one variable ($m_{H}$) fits, with a fixed and floating Higgs boson
  mass, respectively. The solid squares linked with solid and dashed
  lines correspond to the values of significance using three
  kinematics variables fits and all event classifications, with a
  fixed and floating Higgs boson mass, respectively. The solid squares
  linked with solid and dashed lines correspond to the maximum
  sensitivity that can be attained with a combined analysis, which
  enhance the significance by at least $40\%$ with respect to the inclusive case.

\section{Conclusion}
The feasibility of the search for the Standard Model Higgs boson in the
$H\rightarrow \gamma \gamma$ decay with the ATLAS detector at the LHC has been
evaluated. The main experimental aspects of the analysis have been
evaluated using the updated full detector simulation. The 5$\sigma$
signal significance are expected to be accessed with integrated luminosity
$20-30$ fb$^{-1}$.

\fixspacing

\end{document}